# A POLYNOMIAL DISTRIBUTION APPLIED TO INCOME AND WEALTH DISTRIBUTION

Elvis Oltean, Fedor Kusmartsev

e-mail: elvis.oltean@alumni.lboro.ac.uk

Income and wealth distribution affect stability of a society to a large extent and high inequality affects it negatively. Moreover, in the case of developed countries, recently has been proven that inequality is closely related to all negative phenomena affecting society. So far, Econophysics papers tried to analyse income and wealth distribution by employing distributions such as Fermi-Dirac, Bose-Einstein, Maxwell-Boltzmann, lognormal (Gibrat), and exponential. Generally, distributions describe mostly income and less wealth distribution for low and middle income segment of population, which accounts about 90% of the population.

Our approach is based on a totally new distribution, not used so far in the literature regarding income and wealth distribution. Using cumulative distribution method, we find that polynomial functions, regardless of their degree (first, second, or higher), can describe with very high accuracy both income and wealth distribution. Moreover, we find that polynomial functions describe income and wealth distribution for entire population including upper income segment for which traditionally Pareto distribution is used.

Wealth and income distribution are one of the most important issues in a society considering that an optimal level ensures social stability while a high degree causes multiple problems.

A recent study by Wilkinson [1] shows that for the developed countries there is a direct relation between economic inequality and all the social problems that impact on society such as criminality, social trust, obesity, infant mortality, violence, child poverty, mental illness, imprisonment, and many others that characterise the quality of life. Thus, the countries with the lowest inequality, such as Scandinavian countries and Japan, have the best indicators regarding the social phenomena that affect the social life. The opposite is represented by the USA, which has the highest inequality among developed countries. The USA is characterised by the highest impact of negative phenomena affecting negatively the society.

In the more recent years, a distinct field of Econophysics emerged. This one deals with size of firms, macroeconomic aggregates, income and wealth distribution [2], [3].

Traditionally, in the Economics literature Gini coefficient is the most used measure of inequality. More recently, one of the areas of research dedicated to wealth and income distribution belongs to Econophysics. So far, the most used distributions were Bose-Einstein, lognormal (Gibrat), and Fermi-Dirac distributions.

## 1. Short History and Theoretical Background

The modern approach of income and wealth distribution appeared first in the articles of Yakovenko [4], [5], [6], and [7]. Main findings of his work are about Bose-Einstein distribution, lognormal distribution, and exponential distribution which explain the income and wealth distribution for low and middle tier income population.

Kusmartsev, while exploring possible applications of Bose-Einstein distribution in the income distribution, tries to analyse possible correlations with other statistical physics variables such as chemical potential and activity coefficient [8], [9].

While most of papers and distributions so far claim to cover the income and wealth distribution only for low and middle income part of population, there two exceptions. Fermi-Dirac distribution [10] and Tsallis distribution [11] claim to be robust enough in order to explain income distribution for the entire range of income, including for the upper income segment of population which traditionally is described by a Pareto distribution.

## 2. Methodology

The data is about nominal or real income and wealth of the population, where income or wealth of each person/household is ranked in increasing order. Next a segmentation of the population is performed by dividing the entire range of income in ten segments. Each part/segment of the population represents 10 % of the total population. Thus, the lowest income segment is considered to be the first one, while the highest income one is the tenth. These segments are known in the relevant literature as deciles.

In our case, we consider data from several countries such as France, Finland, Romania, and Italy. For each country there are certain particularities.

First, for each segment/part of population, it can be calculated a mean income which is total income earned by all individuals/households divided to the total number of individuals/households. Also, for each segment can be ascertained upper limit on income which represents the individual or household with the highest income of all included in that income segment. The latter term was coined by the National Institute of Statistics from Finland.

Second, is about kind of income considered analysed. For all the countries, data is about net (or disposable) income which is the income that remains available after paying taxes and (if the case) benefits received by individual or household. In case of France, we have other data which describe the entire population or certain socio-economic categories by means of gross income (income before redistribution as National Institute of Statistics from France nominates it), net income (for active people and pensioners), and wealth.

Third, in the case of France and Finland, the data are about real income which means that nominal income is adjusted with Consumer Price Index (CPI) which correlates the evolution of nominal income with the inflation rate. For the other countries, we deal with nominal income.

Fourth, the basic element in the analysis of income and wealth is individual (in case of France, Finland, and Mexico), while for Italy and Romania the data is about households.

The methodology used in order to calculate the probability density distribution for the allocation of money for each segment of population is
cumulative probability distribution function (cdf).

$$( ) = \int^{\infty} ( ) \qquad (1)$$

where p is the probability density function. P represents the fraction of the population with nominal or real income or wealth greater than x.

Thus, for the first decile (the lowest income decile) P represents the population that has an income higher or equal to the mean income or upper limit on income of the first decile, hence equals 100%. Subsequently, for the highest income the cumulative distribution function is 10%.

Thus, on the x-axis we represent nominal or real income of the population divided in ten deciles, while on the y-axis we represent the cumulative probability function for the probability density distribution of the population having certain income or wealth.

We used for representation normal representation without considering log-log representation for the data.

The distribution that we found to fit best the data from a variety of distribution is second degree polynomial function.

$$y=a*x^2+b*x+c$$

3. Data analysis

According to the data available for each country considered, we have two sets of data for France, Finland, and Italy. Thus, one set is about mean income and the other set is about upper limit on income For Romania and Mexico, we have data only about mean income. All the data mentioned before is about net/disposable income. Lastly, in case of France we will analyse data about gross income (income before redistribution), data about active population and pensioners, and wealth.

3.1 Net (Disposable) income

3.1.1 France

The data available for France is about mean income [12] and upper limit on income [13]. The detailed findings are exhibited in the Appendixes 1 and 2.

The data from France is in euro 2009, which means that for the previous years, the nominal income was converted to real income by using Consumer Price Index relative to the prices from the year 2009.

Using cumulative distribution function, we got very good fitting to the data. Thus, for mean income set of data the coefficient of determination ($R^2$) has the lowest value 0.9887 for the year 2009, while the highest value is 0.9918 for the data from the year 2003. In the case of upper limit on income, the coefficient of determination has values above 0.99 for all the years considered.

We can notice that the values for the coefficient of determination are higher in the case of the upper limit on income data set.

In Figure 1, we present an example in order to observe graphically the goodness of the fit to the data.

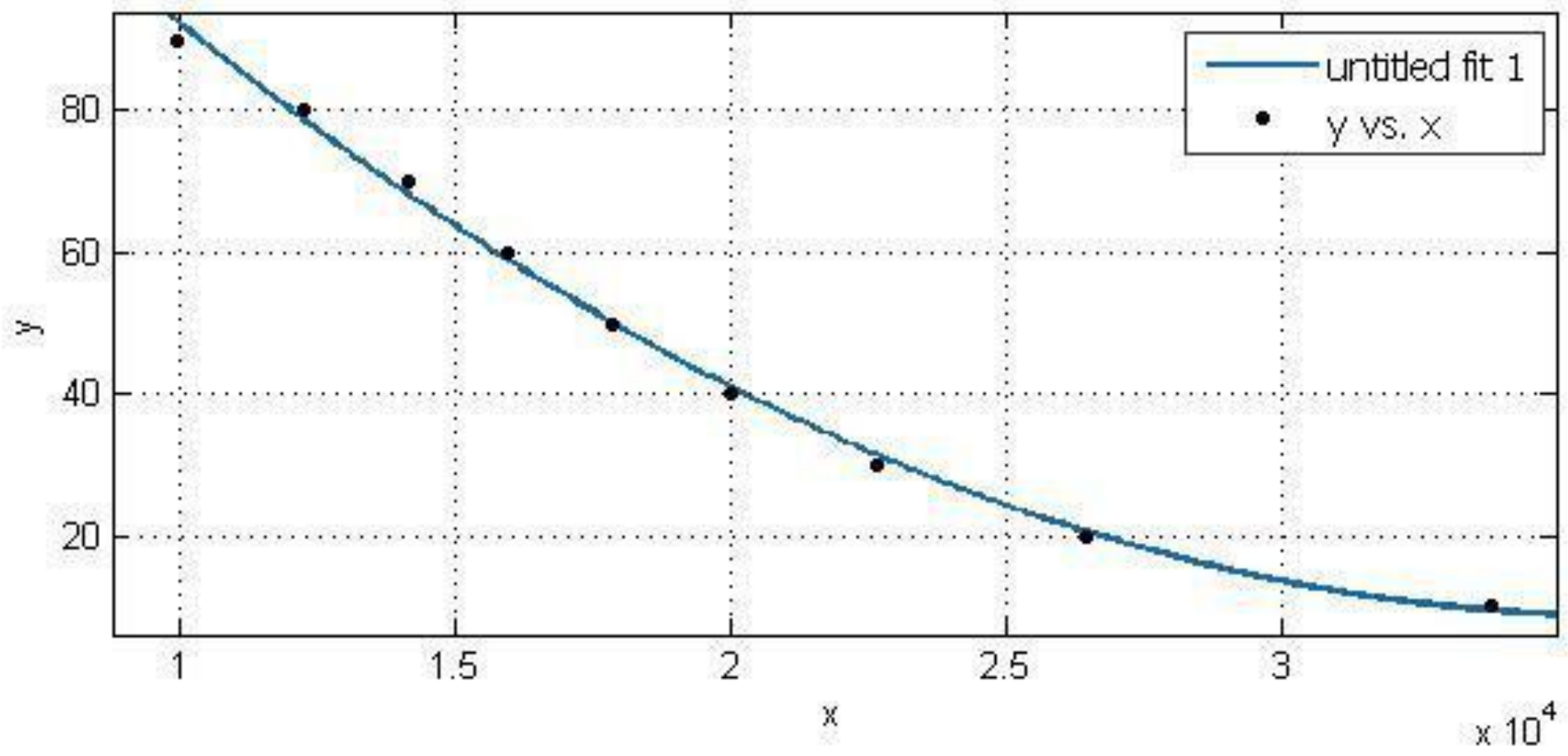

**Figure 1.** Cumulative distribution probability for upper limit on income in the year 2002

The equation describing the distribution is

$$Y = 1.196*10^{-7} + (-0.008721)*x + 167.5, \text{ where } R^2 = 99.72\%$$

3.1.2 Finland

The data about income is in the data sets [14] for mean income and [15] for upper limit on income. The detailed findings are exhibited in the Appendixes 3 and 4.

The data about Finland is real income expressed in euro to the value from the year 2009, which implies that the values of real income from the previous years has been adjusted by taking into account the CPI as a moderator for inflation rate.

The values resulting from the cumulative distribution function in order to fit the data yielded for the coefficient of determination were very good. Thus, for the mean income data set, the lowest value was 0.9758 (corresponding to the year 1991), while the highest value was 0.9914 corresponding to the year 2006. For the upper limit on income data set, the values for the coefficient of determination were above 0.99 in all cases. We can notice that the values of the coefficient of determination were higher for upper limit on income than in the case of mean income data set.

In order to illustrate with an example, we chose the year that presented the lowest value for the coefficient of determination from the upper limit on income data set (1988).

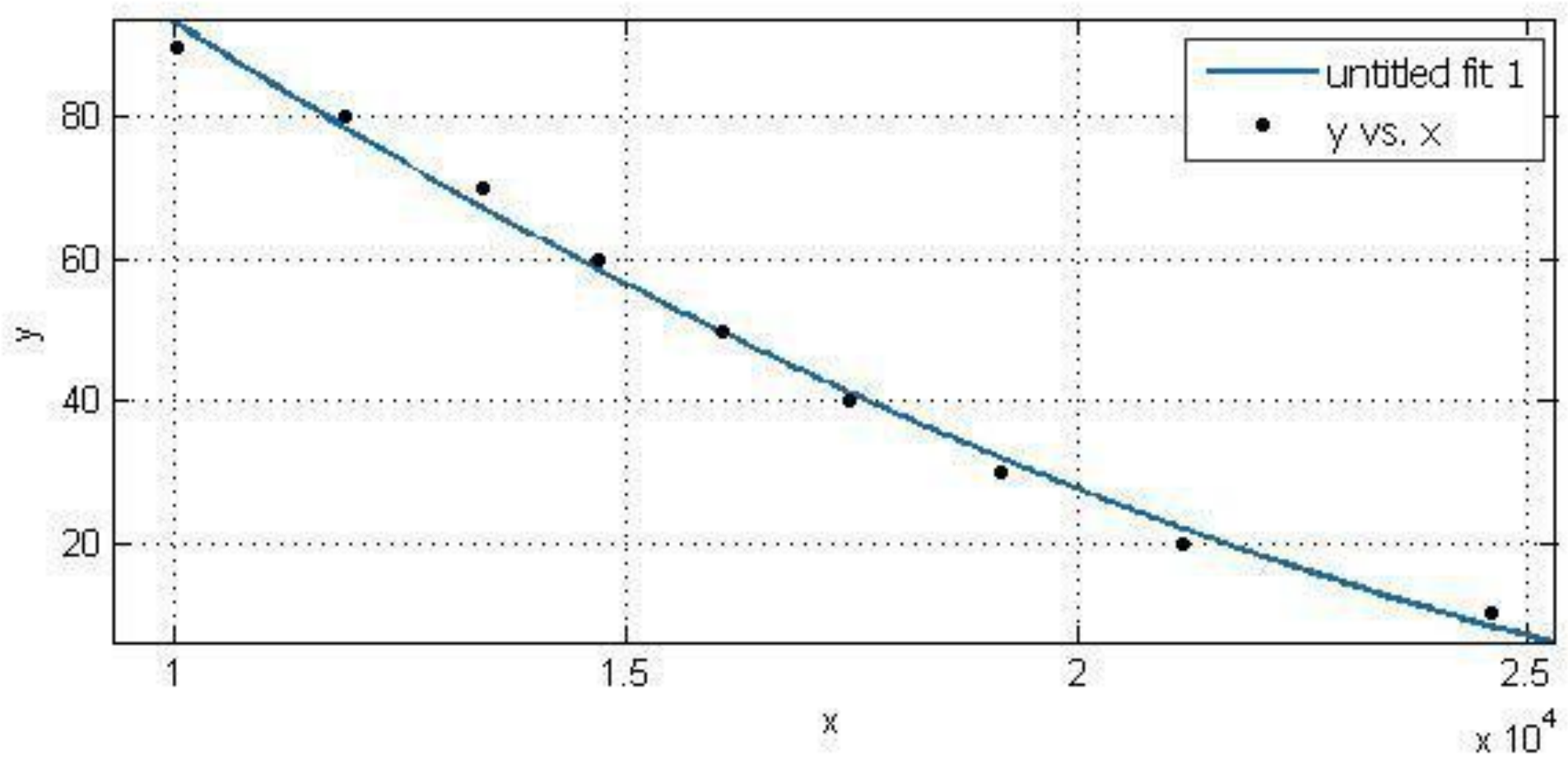

**Figure 2.** Cumulative distribution probability for upper limit on income in the year 1988

The equation describing the probability distribution is

Y= $1.596*10^{-7}$+ (-0.01135)*x+ 190.8, where $R^2$=99.43 %

3.1.3 Romania

The data set about Romania contains information about nominal income using mean income as tool to measure it [16]. The detailed findings are exhibited in the Appendix 5.

Unlike previous cases, we are dealing here with two different currencies. Thus, up to the year 2005, the national currency of Romania was leu. After middle of the year 2005, the new currency introduced was called heavy leu and 1 heavy leu=10000 leu. A characteristic before time interval considered before 2005 was inflation which increased up to 30%.

Cumulative distribution function used to fit the data yielded high values for the coefficient of determination. The lowest value was for the year 2004 (0.975), while the highest value was 0.9958 for the year 2007.

In order to have a graphical image about the goodness of the fit, we present the function fitting the data from the year 2004 in Figure 3.

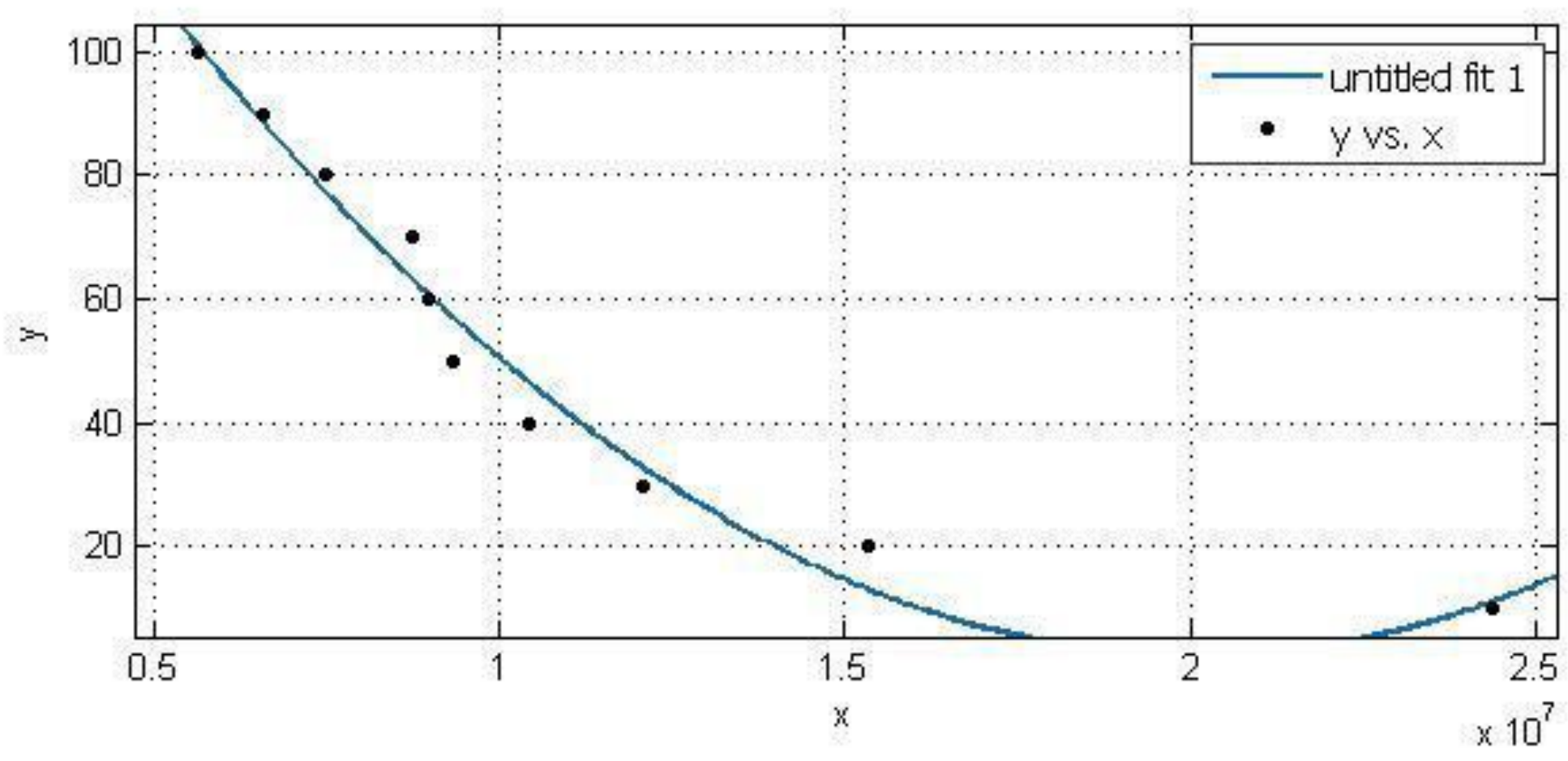

**Figure 3.** Cumulative distribution probability for mean income in the year 2004

The equation describing the distribution is

$$Y= 4.693*10^{-13}+(-1.886*10^{-5})*x+ 191.9, \text{ where } R^2=97.5\ \%$$

We notice that the yearly values from fitting the data using second degree polynomial cdf that there are no significant differences regarding the coefficient of determination regarding the two time intervals corresponding to the two currencies. The only plausible observation regarding the values for the coefficient of determination after the introduction of the new currency, which was a time of high economic growth up to the year 2008 (when the crisis started in Romania) are higher than 0.99.

Regarding the values the parameters, we can notice that their values are significantly lower for the time interval before the year 2005. This is explainable by the higher existing prices and higher inflation in the time interval before the year 2005.

### 3.1.4 Italy

The data sets describing income distribution about mean income and upper limit on income [17] and [18]. The detailed findings are exhibited in the Appendixes 6 and 7. However, just like in the case of Romania, we are dealing with two different currencies. Thus, for the time interval 1989-1998 is expressed in Italian lira, while for the time interval 2000-2008 the data is expressed in euro, nominal values. Italian Lira was a currency characterised by high inflation though the inflation in the 90s was not as high as in the previous decade. Euro can be considered a very stable currency compared to Italian lira.

The second degree polynomial cdf fits the data very well. Thus, both for mean income data set and for upper limit on income values for the coefficient of determination are above 99 %.

In order to illustrate these evolutions, we chose the year 2000 from mean income data set which is exhibited in Figure 4.

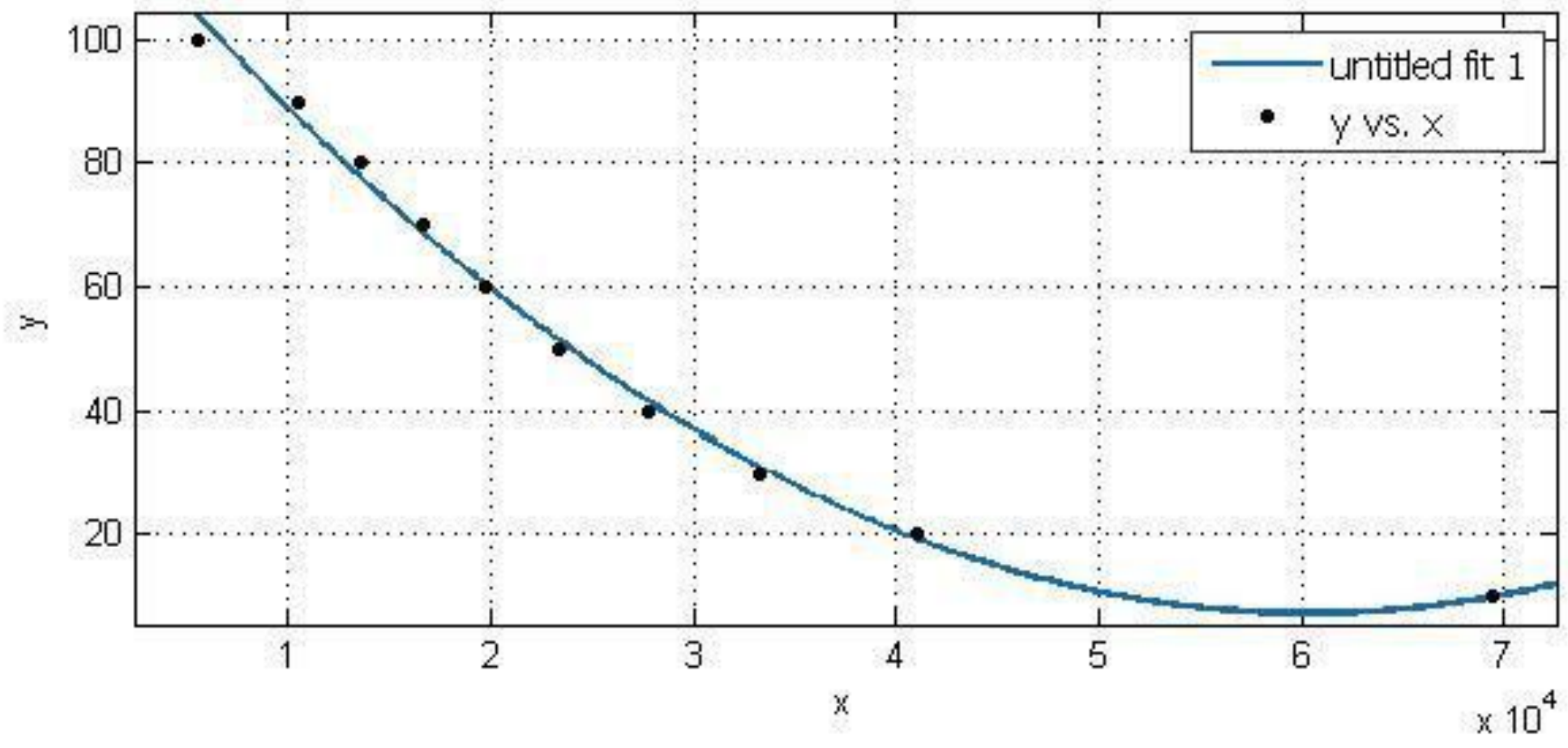

**Figure 4.** Second degree cdf probability for mean income in the year 2004

The equation describing the distribution is

$$Y= 3.213*10^{-8}+(-0.00388 )*x+ 124.4 \text{ , where } R^2=99.58\%$$

We can notice that for both types of data sets (mean income and upper limit on income) the values for the coefficient of the determination are above 0.99 which are the highest of all data sets for the countries considered in the present paper.

We can observe that similarly to Romania, the values for the parameters of the cdf are significantly lower for the time interval when Italian lira was the official currency. Moreover, this is valid for both types of data sets (mean income and upper limit on income)

3.2 Other types of income

Unfortunately, the only data regarding other types of income on one hand and wealth on the other hand were from France. In the following, we present our findings regarding income before redistribution and income of pensioners.

### 3.2.1 Income before redistribution

Income before redistribution or gross income is the income that an individual or a household gets before paying taxes and receiving all kind of social benefits (if the case). The detailed findings are exhibited in the Appendix 8. In the data set from France which spans over time interval 2003-2009 [19], we used to calculate the income distribution upper limit on income. The values of the parameters of the fitting function are not far outside from most of the other fitting functions across our findings. The values for the coefficient of determination are higher than 99 % in all cases.

### 3.2.2 Income of pensioners

We applied second degree polynomial cdf not to the entire population, but to certain social category. The detailed findings are exhibited in the Appendix 9. The data regarding this social category [20] used upper limit on income as method to calculate income distribution. The values regarding the values of the parameters are within the same normal limit discovered among the other distribution, while for the coefficient of determination is higher than 99 %. It is highly significant that there are no major differences compared to situation when this distribution is applied to the entire population. A possible explanation is that the revenues of the pensioners are closely linked to the revenues of the active population.

### 3.2.3 Wealth

For the data set about wealth [21] we used mean wealth in order to represent the wealth distribution. The detailed findings are exhibited in the Appendix 10. In the case of the values of the parameters of the function no special values outside the normal values were reported. However, the values for the coefficient of determination were slightly lower being approximately 96 %.

In order to illustrate that, we present the case for the year 2010 in Figure 5.

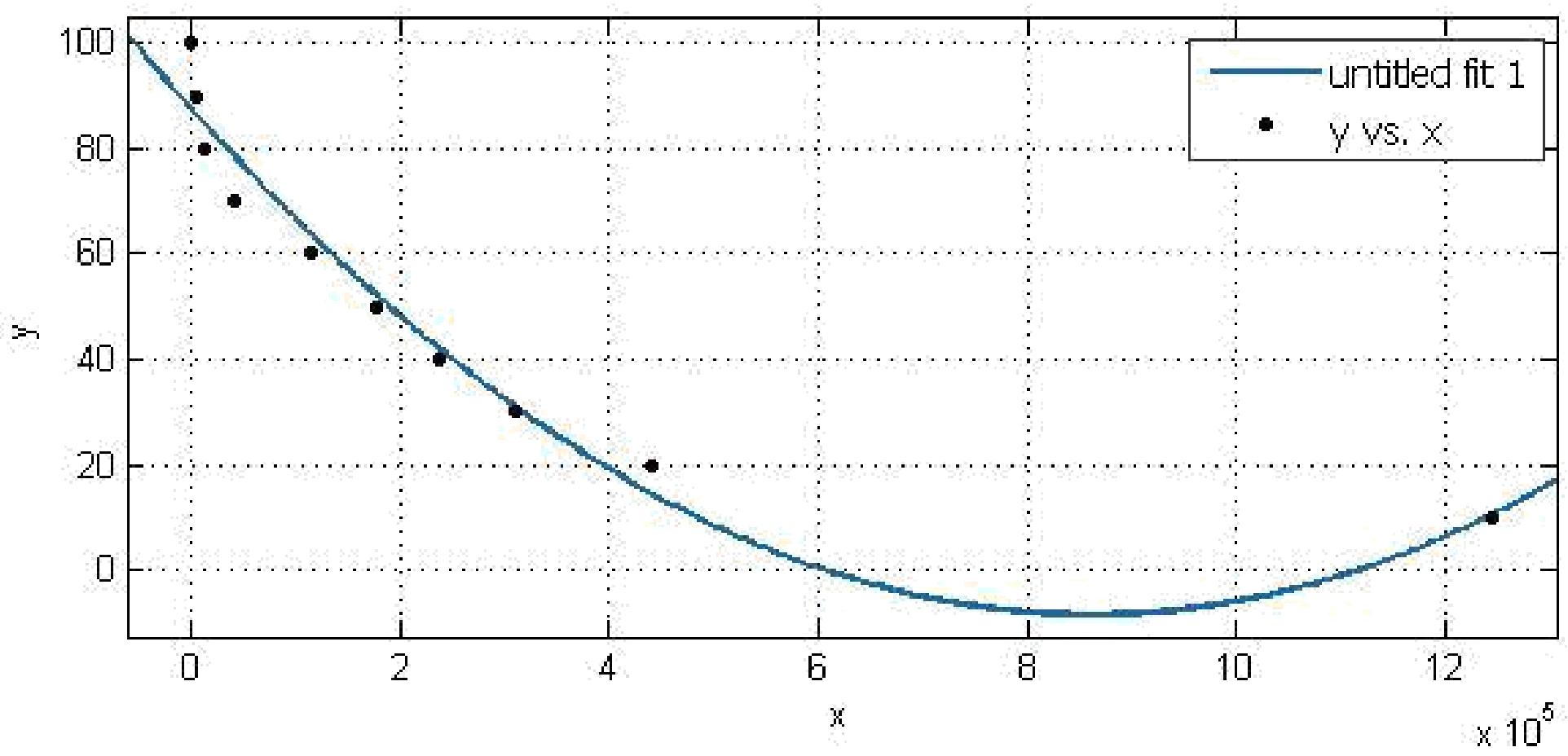

**Figure 5.** Second degree cdf probability for mean wealth in the year 2010

The equation describing the probability distribution is

Y= $1.294*10^{-10}$+(-0.000223 )*x+ 87.6 , where $R^2$=96.15 %

A possible explanation for slightly lower values for the coefficient of determination for all annual probability distributions is that unlike income which depends almost entirely on wages, wealth distribution depends on large extent on asset prices.

4. Results

The values of the coefficient of determination for all mean income data sets have the lowest value around 97 % (in case of Romania), which means that the probability distribution functions fit very well the data.

In case of the upper limit on income data sets, the value regarding the coefficient of determination is in all cases above 0.99, which is almost perfect and better than in the case of mean income data sets. A possible explanation for this is due do the fact that the upper income population (which is comprised in the tenth decile). Thus, upper income population depends solely on income but on asses prices as well, so the underlying mechanism for this segment of population works differently than in the case of the low and middle income part.

While the values of the parameters of the fitting function look generally similar with no relative great discrepancies. There is only one exception in case of Romania, for time interval before the year 2005 and for Italy for time interval before the year 2000 (in both years marked by the introduction of a new and more stable currency). In the cases of these currencies and time intervals, the values of the parameters of the fitting function were significantly lower. This could be explained by the fact wages and prices are high in nominal terms, and inflation was relatively high.

We performed the same analysis using a log-log scale for the values on both x and y axes. Surprisingly, the results were very similar.

In the case of Italy for time interval before the year 2000 (introduction of euro) and Romania for time interval before the year 2005 (introduction of heavy leu), the values for the coefficient of determination were very high. Thus, for Italy the values (both for mean income and upper limit on income data sets) were higher than 99 %, while for Romania the lowest value was about 97%. Thus, we can say that goodness of the fit did not change significantly for currencies and time interval marked by inflation. Moreover, if we compare the same values for the coefficient of determination for Italy both for Italian Lira and Euro, no significant differences can be observed.

For Romania, we found that the values for the coefficient of determination were slightly lower (with few decimals or percent) than for the data from the other countries considered. Considering that for Romania for time interval from the year 2005 onwards where a stable currency was introduced, this could be attributable to the reliability of data. This may be attributable to the high share of the black market.

## 5. Conclusions

We can draw the conclusion that second degree polynomial cumulative distribution function fits well the data we have available. Subsequently, this function is robust and can describe with fairly high accuracy the wealth and income distribution for the population of a country. Moreover, the different economic characteristics of the countries considered (such as different level of development, different macroeconomic characteristics, data reliability) can be considered as an additional proof regarding the robustness of the distribution function.

Given the high values obtained for the coefficient of determination, we can conclude that the second degree polynomial distribution function (as well as higher degree ones) can be used with a high degree of success to explain income and wealth distribution for the entire range of income among population. Thus, both low and middle income part of population and upper income part that traditionally were explained by different distribution functions, can be explained by a single polynomial function.

Polynomial distribution function confirms the main findings of the analysis performed on the same set of data by using Fermi-Dirac distribution regarding income distribution [10].

Upper limit on income data can be used more successfully to explain the income and wealth distribution of income and wealth distribution than mean income data. However, if the analysis aims to look at the entire income range, mean income data is more suitable.

Polynomial distribution function is not affected by inflation, so slightly lower values for the coefficient of determination from the annual analyses may indicate a lower degree of credibility regarding data.

Even though higher degree polynomials can successfully perform the same analysis, we considered that the second degree polynomial is the optimal choice. A higher degree polynomial used as a distribution function would have not significantly improved the accuracy of the analysis and the number of coefficients would have increased.

It is very important that polynomial distribution function can describe successfully both wealth and income distribution. Thus, most of the distributions can solely explain income distribution.

Reference:

[1] Richard Wilkinson, K.Pickett: The Spirit Level: Why More Equal Societies Almost Always Do BetterAllen Lane, 2009.

[2] Burda, Z.; Jurkiewicz, J.; Nowak M.A.; Is Econophysics a Solid Science? www. arXiv:cond-mat/ 0301096 v1, 08.01.2003.

[3] De Liso, N; Filatrella, G.; Econophysics: The emergence of a new field? URL: www.unifr.ch

[4] Adrian A. Dragulescu and Victor M. Yakovenko, Statistical Mechanics of Money, Income, and Wealth: A Short Survey, http://www2.physics.umd.edu/~yakovenk/econophysics.html

[5] A. Dragulescu and V.M. Yakovenko, Statistical mechanics of money, Eur. Phys. J. B **17**, 723{729 (2000)

[6] A.Dragulescu and V.M. Yakovenko, Evidence for the exponential distribution of income in the USA**,** Eur. Phys. J. B **20**, 585{589 (2001)

[7] A. Christian Silva and Victor M. Yakovenko, Temporal evolution of the “thermal” and “superthermal” income classes in the USA during 1983–2001, *Europhys. Lett.*, **69** (2), pp. 304–310 (2005)

[8] K. E. Kurten and F. V. Kusmartsev: Bose-Einstein distribution of money in a free-market economy.Physics Letter A Journal, EPL, 93 (2011) 28003

[9] F. V. Kusmartsev: Statistical mechanics of economics, Physics Letters A 375 (2011) 966– 973

[10] E. Oltean& F. V. Kusmartsev: A study of methods from statistical mechanics to income distribution, ENEC, 2012

[11] J. C. Ferrero, *The equilibrium and nonequilibrium distribution of money, International Workshop on "Econophysics of Wealth Distributions*", Saha Institute of Nuclear Physics, Kolkata, India, 15-19 March, 2005.

[12] Niveau de vie moyens par décile en France, URL: http://www.insee.fr/fr/themes/tableau.asp?reg_id=0&ref_id=NATnon04249

[13] Distribution des niveaux de vie en France, URL: http://www.insee.fr/fr/themes/tableau.asp?reg_id=0&ref_id=NATnon04247

[14] Mean income by decile group in 1987–2009 in Finland. URL: http://www.stat.fi/til/tjt/2009/tjt_2009_2011-05-20_tau_003_en.html

[15] Finland Upper limit on income by decile group in 1987–2009. URL: http://www.stat.fi/til/tjt/2009/tjt_2009_2011-05-20_tau_005_en.html

[16] Buletin statistic, Institutul National de Statistica, Romania, 1999-2011

[17] Supplementi al Bollettino Statistico - Note metodologiche e informazioni statistiche, Numero 14 – 8 Ottobre 1991, Numero 44 - 14 Luglio 1993, Numero 9 - 10 Febbraio 1995, Numero 14 - 20 Marzo 1997, Banca d'Italia, Centro Stampa, Roma

[18] Supplements to the Statistical Bulletin-Methodological notes and statistical information, Number 22 – 18 April 2000, Banca d'Italia, Centro Stampa, Roma

[19] Distribution et moyenne des niveaux de vie avant redistribution en France, URL: http://www.insee.fr/fr/themes/tableau.asp?reg_id=0&ref_id=NATnon04253

[20] Distribution des niveaux de vie des ménages de retraités et des ménages d'actifs en 2009 URL: http://www.insee.fr/fr/themes/tableau.asp?reg_id=0&ref_id=NATnon04254

[21] Patrimoine moyen par décile en 2010, URL: http://www.insee.fr/fr/themes/tableau.asp?reg_id=0&ref_id=patrmoyendecile

Appendix 1 Coefficients from fitting a second degree polynomial distribution to data regarding mean income from France

| France mean income | P1 | P2 | P3 | $R^2$ (%) |
|---|---|---|---|---|
| 2003 | $8.7510^{-8}$ | -0.007277 | 157.3 | 99.18 |
| 2004 | $8.854*10^{-8}$ | -0.007353 | 158.1 | 99.16 |
| 2005 | $8.224*10^{-8}$ | -0.007001 | 154.5 | 98.83 |
| 2006 | $7.797*10^{-8}$ | -0.006823 | 154 | 98.99 |
| 2007 | $7.531*10^{-8}$ | -0.006695 | 153.9 | 99.01 |
| 2008 | $7.325*10^{-8}$ | -0.006643 | 154.9 | 98.77 |
| 2009 | $7.081*10^{-8}$ | -0.006458 | 152.7 | 98.87 |

Appendix 2 Coefficients from fitting a second degree polynomial distribution to data regarding upper limit on income from France

| France upper limit on income | P1 | P2 | P3 | $R^2$ (%) |
|---|---|---|---|---|
| 2002 | $1.196*10^{-7}$ | -0.008721 | 167.5 | 99.72 |
| 2003 | $1.202*10^{-7}$ | -0.00877 | 168 | 99.65 |
| 2004 | $1.212*10^{-7}$ | -0.008837 | 168.7 | 99.65 |
| 2005 | $1.166*10^{-7}$ | -0.008631 | 167.3 | 99.61 |
| 2006 | $1.132*10^{-7}$ | -0.00848 | 167 | 99.62 |
| 2007 | $1.071e*10^{-7}$ | -0.008239 | 166.2 | 99.66 |
| 2008 | $1.076*10^{-7}$ | -0.008327 | 169 | 99.47 |
| 2009 | $1.007*10^{-7}$ | -0.007953 | 164.9 | 99.58 |

Appendix 3 Coefficients from fitting a second degree polynomial distribution to data regarding mean income from Finland

| Finland mean income | P1 | P2 | P3 | $R^2$ (%) |
|---|---|---|---|---|
| 1987 | $1.542*10^{-7}$ | -0.01046 | 180.6 | 97.68 |
| 1988 | $1.56*10^{-7}$ | -0.0105 | 182.4 | 97.73 |
| 1989 | $1.466*10^{-7}$ | -0.01019 | 184 | 98.06 |
| 1990 | $1.36 *10^{-7}$ | -0.00987 | 185.9 | 98.08 |
| 1991 | $1.3*10^{-7}$ | -0.009634 | 184.8 | 97.58 |
| 1992 | $1.563*10^{-7}$ | -0.01073 | 191.3 | 97.6 |
| 1993 | $1.704*10^{-7}$ | -0.01116 | 191 | 98.07 |
| 1994 | $1.715*10^{-7}$ | -0.01122 | 191.7 | 98.26 |
| 1995 | $1.59*10^{-7}$ | -0.01072 | 188.7 | 98.43 |
| 1996 | $1.646*10^{-7}$ | -0.01113 | 196.1 | 99.53 |
| 1997 | $1.258*10^{-7}$ | -0.009276 | 178.5 | 98.88 |
| 1998 | $1.097*10^{-7}$ | -0.008533 | 173 | 98.95 |
| 1999 | $9.903*10^{-8}$ | -0.008142 | 172 | 99.03 |
| 2000 | $9.233*10^{-8}$ | -0.007807 | 169 | 99.25 |
| 2001 | $8.936e*10^{-8}$ | -0.007604 | 168.1 | 98.87 |
| 2002 | $8.404*10^{-8}$ | -0.007361 | 167.8 | 98.79 |
| 2003 | $7.945 *10^{-8}$ | -0.007168 | 167.8 | 99.00 |
| 2004 | $6.946 *10^{-8}$ | -0.006653 | 164.9 | 98.99 |
| 2005 | $6.515 *10^{-8}$ | -0.006446 | 164.9 | 9903 |
| 2006 | $6.207*10^{-8}$ | -0.00626 | 162.9 | 99.14 |
| 2007 | $5.695*10^{-8}$ | -0.005967 | 160.4 | 99.08 |
| 2008 | $5.919*10^{-8}$ | -0.006083 | 162.2 | 98.72 |
| 2009 | $5.709*10^{-8}$ | -0.005981 | 163.8 | 98.69 |

Appendix 4 Coefficients from fitting a second degree polynomial distribution to data regarding upper limit on income from Finland

| Finland upper limit on income | P1 | P2 | P3 | $R^2$ (%) |
|---|---|---|---|---|
| 1987 | $1.557*10^{-7}$ | -0.01131 | 189.3 | 99.52 |
| 1988 | $1.596*10^{-7}$ | -0.01135 | 190.8 | 99.43 |
| 1989 | $1.393*10^{-7}$ | -0.01054 | 187.8 | 99.47 |
| 1990 | $1.478*10^{-7}$ | -0.01089 | 195.8 | 99.49 |
| 1991 | $1.459*10^{-7}$ | -0.01092 | 198.4 | 99.47 |
| 1992 | $1.752*10^{-7}$ | -0.01213 | 205.3 | 99.47 |
| 1993 | $2.206*10^{-7}$ | -0.01359 | 212.4 | 99.68 |
| 1994 | $2.189*10^{-7}$ | -0.01344 | 210.5 | 99.55 |
| 1995 | $1.966 *10^{-7}$ | -0.01257 | 204.4 | 99.56 |
| 1996 | $1.686*10^{-7}$ | -0.0114 | 194.8 | 99.59 |
| 1997 | $1.589*10^{-7}$ | -0.01091 | 191.9 | 99.81 |
| 1998 | $1.284*10^{-7}$ | -0.009596 | 181.4 | 99.76 |
| 1999 | $1.201*10^{-7}$ | -0.009264 | 180.7 | 99.72 |
| 2000 | $1.141*10^{-7}$ | -0.008902 | 176.9 | 99.69 |
| 2001 | $1.03*10^{-7}$ | -0.008464 | 175 | 99.65 |
| 2002 | $9.522*10^{-8}$ | -0.008124 | 174 | 99.57 |
| 2003 | $9.41*10^{-8}$ | -0.008059 | 175 | 99.65 |
| 2004 | $7.89*10^{-8}$ | -0.007303 | 169.8 | 99.59 |
| 2005 | $7.327*10^{-8}$ | -0.007019 | 168.9 | 99.51 |
| 2006 | $7.328*10^{-8}$ | -0.006974 | 168.3 | 99.62 |
| 2007 | $6.924*10^{-8}$ | -0.006729 | 166.6 | 99.57 |
| 2008 | $6.561*10^{-8}$ | -0.006629 | 166.7 | 99.46 |
| 2009 | $6.105*10^{-8}$ | -0.006416 | 167.3 | 99.46 |

Appendix 5 Coefficients from fitting a second degree polynomial distribution to data regarding mean income from Romania

| Romania mean income | P1 | P2 | P3 | $R^2$ (%) |
|---|---|---|---|---|
| 2000 | $4.656 *10^{-12}$ | $-5.821*10^{-5}$ | 187.7 | 98.95 |
| 2001 | $2.31*10^{-12}$ | $-4.332*10^{-5}$ | 205.2 | 98.37 |
| 2002 | $1.3*10^{-12}$ | $-3.134*10^{-5}$ | 193.2 | 98.35 |
| 2003 | $9.795*10^{-13}$ | $-2.799*10^{-5}$ | 202.8 | 97.76 |
| 2004 | $4.693*10^{-13}$ | $-1.886*10^{-5}$ | 191.9 | 97.50 |
| 2005 | $3.233*10^{-5}$ | -0.1491 | 176 | 98.85 |
| 2006 | $2.187*10^{-5}$ | -0.1188 | 165.7 | 99.22 |
| 2007 | $1.541*10^{-5}$ | -0.1006 | 169.4 | 99.58 |
| 2008 | $1.04*10^{-5}$ | -0.08399 | 175.8 | 99.41 |
| 2009 | $9.899 *10^{-6}$ | -0.08451 | 187 | 98.98 |
| 2010 | $1.132*10^{-5}$ | -0.09329 | 199.2 | 98.61 |

Appendix 6 Coefficients from fitting a second degree polynomial distribution to data regarding mean income from Italy

| Italy nominal mean income | P1 | P2 | P3 | $R^2$ (%) |
|---|---|---|---|---|
| 1989 lire | $2.018*10^{-14}$ | $-3.141*10^{-6}$ | 130.3 | 99.73 |
| 1991 lire | $1.773*10^{-14}$ | $-2.944*10^{-6}$ | 131.4 | 99.7 |
| 1993 lire | $1.347*10^{-14}$ | $-2.482*10^{-6}$ | 122.4 | 99.73 |
| 1995 lire | $1.194*10^{-14}$ | $-2.362*10^{-6}$ | 124.2 | 99.73 |
| 1998 lire | $8.948*10^{-15}$ | $-2.03*10^{-6}$ | 121.8 | 99.68 |
| 2000 euro | $3.213*10^{-8}$ | -0.00388 | 124.4 | 99.58 |
| 2002 euro | $2.901*10^{-8}$ | -0.003707 | 125.6 | 99.74 |
| 2004 euro | $2.803*10^{-8}$ | -0.003719 | 129.4 | 99.59 |
| 2006 euro | $2.477*10^{-8}$ | -0.003523 | 131.3 | 99.78 |
| 2008 euro | $2.307*10^{-8}$ | -0.003351 | 128.4 | 99.65 |

Appendix 7 Coefficients from fitting a second degree polynomial distribution to data regarding upper limit on income from Italy

| Italy upper limit on income | P1 | P2 | P3 | $R^2$ (%) |
|---|---|---|---|---|
| 1989 lire | $2.572*10^{-14}$ | $-3.573*10^{-6}$ | 132.6 | 99.95 |
| 1991 lire | $2.001*10^{-14}$ | $-3.165*10^{-6}$ | 131.1 | 99.91 |
| 1993 lire | $1.67*10^{-14}$ | $-2.76*10^{-6}$ | 122.4 | 99.87 |
| 1995 lire | $1.489*10^{-14}$ | $-2.639*10^{-6}$ | 124.7 | 99.9 |
| 1998lire | $1.142*10^{-14}$ | $-2.296*10^{-6}$ | 123 | 99.92 |
| 2000 euro | $4.181*10^{-8}$ | -0.004456 | 126.9 | 99.91 |
| 2002 euro | $3.783*10^{-8}$ | -0.004244 | 127.7 | 99.97 |
| 2004 euro | $3.884*10^{-8}$ | -0.004378 | 133 | 99.88 |
| 2006 euro | $3.296*10^{-8}$ | -0.004069 | 134.2 | 99.96 |
| 2008 euro | $3.105*10^{-8}$ | -0.00389 | 131.3 | 99.88 |

Appendix 8 Coefficients from fitting a second degree polynomial distribution to data regarding upper limit on income before distribution from France

| France upper limit on income before distribution | P1 | P2 | P3 | $R^2$ (%) |
|---|---|---|---|---|
| 2003 | $6.534 * 10^{-8}$ | -0.005766 | 130.3 | 99.36 |
| 2004 | $6.644 * 10^{-8}$ | -0.005849 | 131.3 | 99.25 |
| 2005 | $6.466 * 10^{-8}$ | -0.005755 | 131.1 | 99.24 |
| 2006 | $6.465 * 10^{-8}$ | -0.005743 | 131.4 | 99.26 |
| 2007 | $5.991 * 10^{-8}$ | -0.005535 | 130.3 | 99.29 |
| 2008 | $6.039 * 10^{-8}$ | -0.005569 | 131.8 | 99.06 |
| 2009 | $5.541 * 10^{-8}$ | -0.005312 | 129.3 | 99.16 |

Appendix 9 Coefficients from fitting a second degree polynomial distribution to data regarding upper limit on income of pensioners from France

| France upper limit on income of pensioners | P1 | P2 | P3 | $R^2$ (%) |
|---|---|---|---|---|
| 2002 | $1.531 * 10^{-7}$ | -0.01025 | 181.8 | 99.88 |
| 2003 | $1.626 * 10^{-7}$ | -0.01069 | 186.2 | 99.88 |
| 2004 | $1.645 * 10^{-7}$ | -0.01082 | 187.9 | 99.85 |
| 2005 | $1.579 * 10^{-7}$ | -0.01052 | 185.4 | 99.79 |
| 2006 | $1.377 * 10^{-7}$ | -0.009655 | 179.5 | 99.83 |
| 2007 | $1.39 * 10^{-7}$ | -0.009761 | 181.4 | 99.83 |
| 2008 | $1.377 * 10^{-7}$ | -0.00975 | 182.6 | 99.73 |
| 2009 | $1.333 * 10^{-7}$ | -0.009555 | 181.2 | 99.87 |

Appendix 10 Coefficients from fitting a second degree polynomial distribution to data regarding mean wealth from France

| France mean wealth | P1 | P2 | P3 | $R^2$ (%) |
|---|---|---|---|---|
| 1998 | $6.227 * 10^{-10}$ | -0.0004848 | 88.4 | 96.7 |
| 2004 | $3.181 * 10^{-10}$ | -0.0003423 | 87.7 | 96.36 |
| 2010 | $1.294 * 10^{-10}$ | -0.000223 | 87.6 | 96.15 |